\documentclass[prd,preprint,superscriptaddress,amsmath,amssymb,nofootinbib]{revtex4}
\usepackage{graphicx}% Include figure files
\usepackage{dcolumn}% Align table columns on decimal point
\usepackage{bm}% bold math
\usepackage{amssymb}
\usepackage{amsmath}
\usepackage{epsfig}    
\usepackage{color}
\usepackage{slashed}
\usepackage{hhline}
\def\be{\begin{equation}}
\def\ee{\end{equation}}
\newcommand{\bea}{\begin{eqnarray}}
\newcommand{\eea}{\end{eqnarray}}
\newcommand{\nn}{\nonumber}

\begin{document}

 \begin{flushright} {KIAS-P18115}, APCTP Pre2018 - 017  \end{flushright}

%%%%%%%%%
\title{
%One-loop neutrino mass model with $SU(2)_L$ multiplet fields
A one-loop neutrino mass model with $SU(2)_L$ multiplet fields
}

\author{Takaaki Nomura}
\email{nomura@kias.re.kr}
\affiliation{School of Physics, KIAS, Seoul 02455, Republic of Korea}

\author{Hiroshi Okada}
\email{hiroshi.okada@apctp.org}
\affiliation{Asia Pacific Center for Theoretical Physics, Pohang, Gyeongbuk 790-784, Republic of Korea}

\date{\today}

\begin{abstract}
%
%We propose a one-loop neutrino mass model with several $SU(2)_L$ multiplet fermions and scalar fields in which the inert feature of a scalar to realize the one-loop neutrino mass can be achieved by the cancellation among Higgs couplings thanks to nontrivial terms in the Higgs potential. 
{We propose a one-loop neutrino mass model with several $SU(2)_L$ multiplet fermions and scalar fields in which the inert feature of a scalar to realize the one-loop neutrino mass can be achieved by the cancellation among Higgs couplings thanks to non-trivial terms in the Higgs potential and to present it in a simpler way.}
Then we discuss our typical cut-off scale by computing renormalization group equation for $SU(2)_L$ gauge coupling, lepton flavor violations, muon anomalous magnetic moment,  possibility of dark matter candidate, neutrino mass matrix satisfying the neutrino oscillation data. Finally, we search for our allowed parameter region to satisfy all the constraints, and discuss a possibility of detecting new charged particles at the large hadron collider. 
 \end{abstract}
\maketitle

\section{Introductions}

Radiatively induced neutrino mass models are one of the promising candidates to realize tiny neutrino masses with natural parameter spaces at TeV scale and to provide a dark matter (DM) candidate, both of which cannot be explained within the standard model (SM).
In order to build such a radiative model, an inert scalar boson plays an important role and its inert feature can frequently be realized by imposing additional symmetry such as  $Z_2$ symmetry~\cite{Ma:2006km, Krauss:2002px, Aoki:2008av, Gustafsson:2012vj} and/or $U(1)$ symmetry~\cite{Okada:2012np, Kajiyama:2013zla, Kajiyama:2013rla}, which also play an role in stabilizing the DM. 
On the other hand, once we introduce large $SU(2)_L$ multiplet fields such as quartet~\cite{Nomura:2018ktz, Nomura:2018ibs}, quintet~\cite{Nomura:2018lsx, Nomura:2018cle}, septet fields~\cite{Nomura:2018cfu, Nomura:2017abu, Nomura:2016jnl}, we sometimes can evade imposing additional symmetries~\cite{Anamiati:2018cuq, Cirelli:2005uq}. Then, the stability originates from a remnant symmetry after the spontaneous electroweak symmetry breaking due to the largeness of these multiplets. In addition, the cut-off scale of a model is determined by the renormalization group equations (RGEs) of $SU(2)_L$ gauge coupling, and it implies that  a theory  can be within TeV scale, depending on the number of multiplet fields. Thus a good testability could be provided in such a scenario.

{
Then, using large $SU(2)_L$ multiplet fields, we would like to realize one-loop neutrino generation by inert scalar field without imposing additional symmetry such as $Z_2$.  
In this case scalar quintet $H_5$ is minimal choice for inert multiplet since scalar multiplet smaller than quintet easily develop a vacuum expectation value (VEV) by renormalizable interaction 
with SM Higgs field $H$ like $H_4 H H H$ for the quadruplet $H_4$.
In addition we need quadruplet fermion $\psi_4$ to interact $H_5$ with the SM lepton doublet and septet scalar $H_7$ is also required to get Majorana mass term from $\psi_4$ by its VEV (Higgs triplet is also possible but it allows type-II seesaw mechanism~\cite{Magg:1980ut, Konetschny:1977bn}).
We find that scalar quadruplet $H_4$ is needed to realize vacuum configuration in which the VEV of $H_5$ to be zero; in addition we can avoid dangerous massless Goldstone boson from scalar multiplets by non-trivial terms with these multiplets. 
Although number of exotic fields is smaller in other one-loop neutrino mass models like scotogenic model~\cite{Ma:2006km} they usually require additional discrete symmetry such as $Z_2$.
We show the realization of one-loop neutrino mass without additional symmetry which result in introduction of several exotic multiplets.
}

In this letter, we introduce several multiplet fermions and scalar fields under the $SU(2)_L$ gauge symmetry. As a direct consequence of multiplet fields, 
our cut-off scale is of the order 10 PeV that could be tested by current or future experiments.
In our model we do not impose additional symmetry and search for possible solution to obtain inert condition for generating neutrino mass at loop level.
Then required inert feature can be realized not via a remnant symmetry   
but via cancellations among couplings in our scalar potential thanks to several non-trivial couplings~\cite{Okada:2015bxa}. 
In such a case, generally DM could decay into the SM particles, but we can control some parameters so that we can evade its too short lifetime without requiring too small couplings. 
Therefore our DM is long-lived particle which represents clear difference from the scenario where the stability of DM is due to an additional or remnant symmetry.
We also discuss lepton flavor violations (LFVs), and anomalous magnetic moment (muon $g-2$), and search for allowed parameter region to satisfy all the constraints such as neutrino oscillation data, LFVs, DM relic density, and demonstrate the possibility of detecting new charged particles at the large hadron collider (LHC).

This letter is organized as follows.
In Sec.~II, {we review our model and formulate the Higgs sector,  neutral fermion sector including active neutrinos.
Then we discuss the RGE of the $SU(2)_L$ gauge coupling, LFVs, muon $g-2$, and our DM candidate.
In Sec.~III, we explore the allowed region to satisfy all the constraints, and discuss production of our new fields (especially charged bosons) at he LHC. 
In Sec.~IV, we devote the summary of our results and the conclusion.}

\section{Model setup and Constraints}
\begin{table}[t!]
\begin{tabular}{|c||c|c|c||c|c|c|c|}\hline\hline  
& ~$L_L^a$~& ~$e_R^a$~& ~$\psi^a$~& ~$H_2$~& ~$H_4$~& ~$H_5$~& ~$H_7$~\\\hline\hline 
%%%
$SU(2)_L$   & $\bm{2}$  & $\bm{1}$  & $\bm{4}$  & $\bm{2}$ & $\bm{4}$  & $\bm{5}$    & $\bm{7}$   \\\hline 
$U(1)_Y$    & -$\frac12$  & -$1$ & {-$\frac12$}  & $\frac12$  & $\frac12$   &{$0$}  &{$1$} \\\hline
\end{tabular}
\caption{Charge assignments of the our lepton and scalar fields
under $SU(2)_L\times U(1)_Y$, where the upper index $a$ is the number of family that runs over 1-3 and
all of them are singlet under $SU(3)_C$. }\label{tab:1}
\end{table}

In this section we formulate our model.
As for the fermion sector, we introduce three families of vector-like fermions $\psi$ with $(4,{-1/2})$ charge under the $SU(2)_L\times U(1)_Y$ gauge symmetry.
As for the scalar sector, we respectively add an $SU(2)_L$ quartet ($H_4$), quintet ($H_5$), and septet ($H_7$) complex scalar fields  with {$(1/2,0,1)$} charge under the $U(1)_Y$ gauge symmetry in addition to the SM-like Higgs that is denoted by $H_2$, where the quintet $H_5$ is expected to be an inert scalar field.
Here we write the nonzero vacuum expectation values {(VEVs)} of $H_2$, $H_4$, and $H_7$ by $\langle H_2\rangle\equiv v_H/\sqrt2$, $\langle H_4\rangle\equiv v_4/\sqrt2$ and $\langle H_7\rangle\equiv v_7/\sqrt2$, respectively, which induces the spontaneous electroweak symmetry breaking.
All the field contents and their assignments are summarized in Table~\ref{tab:1}, where the quark sector is exactly the same as the SM.
%%%
The renormalizable Yukawa Lagrangian under these symmetries is given by
\begin{align}
-{\cal L_\ell}
& =  y_{\ell_{aa}} \bar L^a_L H_2 e^a_R  +  f_{ab} [ \bar L^a_L H_5 (\psi_R)^b ]  
 +  g_{L_{aa}} [(\bar\psi^c_L)^a  H_7 \psi^a_L]
+ g_{R_{aa}} [(\bar\psi^c_R)^a  H_7\psi^a_R] \nn\\
&+M_{D_{aa}} \bar \psi^a_R \psi_L^a + {\rm h.c.}, \label{Eq:yuk}
% \\
%%%& {\cal V}= -\mu^2_1 |\varphi|^2 -\mu^2_2 |H|^2 + \lambda_1 |\varphi|^4 + \lambda_2 |H|^4  + \lambda_{3} |\varphi|^2 |H|^2 ,\label{Eq:pot}
\end{align}
where $SU(2)_L$ index is omitted assuming it is contracted to be gauge invariant inside bracket [$\cdots$], 
upper indices $(a,b)=1$-$3$ are the number of families, and $y_\ell$ and either of $g_{L/R}$ or $M_D$ are assumed to be  diagonal matrix with real parameters without loss of generality. Here, we assume $g_{L/R}$ and $M_D$ to be diagonal for simplicity.
The mass matrix of charged-lepton is defined by $m_\ell=y_\ell v/\sqrt2$. 
Here we assign lepton number $1$ to $\psi_{}$ so that the source of lepton number violation is only the terms with coupling $g_{ab}$ and $g'_{ab}$ in the Lagrangian requiring the lepton number is conserved at high scale.

\subsection{Scalar sector}
\noindent \underline{\it Scalar potential and VEVs}:
The  scalar potential in our model is given by 
{\begin{align}
{\cal V} = &  - M_2^2 H_2^\dagger H_2  + M_4^2 H_4^\dagger H_4 + M_7^2 H_7^\dagger H_7 + \lambda_{H} (H_2^\dagger H_2)^2 \nn \\
&+ \mu_H^2 [H_5^2] + \mu_1 [H_2 \tilde H_4 H_5] + \mu_2 [H_4^T \tilde H_7 H_4]+ \lambda_0 [H_2^T H_2 H_5 H_7^*] \nn\\
& + \lambda_1 [H_2 H_4 H_5 \tilde H_7] + \lambda_2 [H_2^\dag H_2 H_4^\dag H_2]  +{\rm h.c.} + V_{tri},
%+ {\cal V}_{\rm trivial}, 
\label{Eq:potential}
\end{align}
where $V_{tri}$ is the trivial quartic terms containing $H_{4,5,7}$.  }
From the conditions of $\partial {\cal V}/\partial v_5 = 0$ and $\langle H_5\rangle=0$, we find the following relation: 
\begin{align}
v_4 =\frac{3\sqrt{10} v_7 v_2 \lambda_0}{\sqrt{30}v_7\lambda_1+15 \mu_1}  \label{eq:cond1}.
\end{align}
{Then, the stable conditions to the $H_4$ and $H_7$ lead to the following equations:
\begin{align}
v_2 = \frac{3}{8} \left( \frac{\lambda_2}{\lambda_H} v_4 + \sqrt{\frac{\lambda_2^2}{\lambda_H^2}v_4^2 + \frac{64 M_2^2}{9 \lambda_H} } \right), \quad
%\frac{4 \Lambda^2_2 }{\sqrt{3} \lambda_2 v_4 }, \quad
v_4 =\frac{5v_2^3 \lambda_2 }{2\sqrt{3}(10 M^2_4 +\sqrt{30} \mu_2)}, \quad
v_7  = -\sqrt{\frac{3}{10}}\frac{v_4^2 \mu_2 }{2 M^2_7}, \label{eq:cond2}
\end{align}
where we have ignored contributions from terms in $V_{tri}$ assuming corresponding couplings are negligibly small; we can always find a solution satisfying the inert condition including such terms.}
Solving Eqs.(\ref{eq:cond1}) and (\ref{eq:cond2}), one rewrites VEVs and  one parameter in terms of the other parameters.  
In addition to the above conditions, we also need to consider the constraint from $\rho$ parameter, which is given by the following relation at tree level:
\begin{align}
\rho\approx \frac{v_2^2+\frac{11}2 v_4^2+22v_7^2}{v_2^2 + v_4^2 + 4v_7^2},
\end{align}
where the experimental values is given by $\rho=1.0004^{+0.0003}_{-0.0004}$ at 2$\sigma$ confidential level~\cite{pdg}.
Then, we have, e.g., the solutions of $(v_2,v_4,v_7)\approx(246, 2.18,1.03)$ GeV, where $v_2^2 + v_4^2 + 4v_7^2\approx 246$ GeV$^2$.

\if0
\noindent \underline{\it Exotic particles} :
The scalars and fermions with large $SU(2)_L$ multiplet provide exotic charged particles.
Here we write components of multiplets as
\begin{align}
& H_5 = (\phi_5^{++}, \phi_5^{+}, \phi_5^{0}, \phi'^{-}_5, \phi'^{--}_5)^T,  \label{eq:H5} \\
& \psi_{L(R)} = (\psi^{+}, \psi^{0}, \psi'^{-}, \psi^{--})^T_{L(R)},  \label{eq:psiLR} \\
& \Sigma_R = (\Sigma^{++}, \Sigma^{+}, \Sigma^{0}, \Sigma'^{-}, \Sigma'^{--})_R^T. \label{eq:sigmaR} 
\end{align}
The mass of component in $H_5$ is given by $\sim M_5$ where charged particles in the same multiplet have degenerate mass at tree level which will be shifted at loop level~\cite{Cirelli:2005uq}. 
For charged fermions, components from $\psi_{L(R)}$ and $\Sigma_R$ can be mixed after electroweak symmetry breaking via Yukawa coupling.
If the Yukawa couplings are negligibly small  
the charged components in $\psi_{L(R)}$ have Dirac mass $M_D$ while
the charged components in $\Sigma_R$ have Dirac mass $M_\Sigma$ where mass terms are constructed by pairs of positive-negative charged components in the multiplet.
Note that mass term of neutral component is discussed with neutrino sector below.
\fi

\if0
\begin{figure}[tb]
\begin{center}
\includegraphics[width=5.0cm]{mu_mass.eps}
\caption{Feynman diagram to generate the masses of $\mu_{L/R}$.}
\label{fig:mu_mass}
\end{center}\end{figure}
\fi

\subsection{Neutral fermion masses}
\noindent \underline{\it Heavier neutral sector}:
After the spontaneously electroweak symmetry breaking, extra neutral fermion mass matrix in basis of $\Psi^0_R\equiv (\psi^0_R,\psi_L^{0c})^T$ is given by
\begin{align}
M_N
&=
\left[\begin{array}{cc}
\mu_R & M_D^T   \\ 
M_D &  \mu_L \\ 
\end{array}\right],
\end{align}
where $\mu_{R}\equiv \sqrt{\frac{3}{10}}g_{R} v_7$ and $\mu_{L}\equiv \sqrt{\frac{3}{10}}g^*_{L} v_7$.
Since we can suppose hierarchy of mass parameters to be $\mu_{L/R}<<M_D$, the mixing is expected to be maximal.
Thus, we formulate the eigenstates in terms of the flavor eigenstate as follows:
\begin{align}
\psi^0_R=\frac{i}{\sqrt2} \psi_{1_R} - \frac{i}{\sqrt2} \psi_{2_L}^{c},\quad
\psi^{0c}_L=\frac{1}{\sqrt2} \psi_{1_R} + \frac{1}{\sqrt2} \psi_{2_L}^{c},
\end{align}
where $\psi_{1_R}$ and $\psi^c_{2_L}$ represent the mass eigenstates, and their masses are respectively given by $M_a\equiv M_D- (\mu_R+\mu_L)/2$ (a=1-3)
 $M_b\equiv M_D + (\mu_R+\mu_L)/2$ (b=4-6).

\begin{figure}[t]
\begin{center}
\includegraphics[width=10cm]{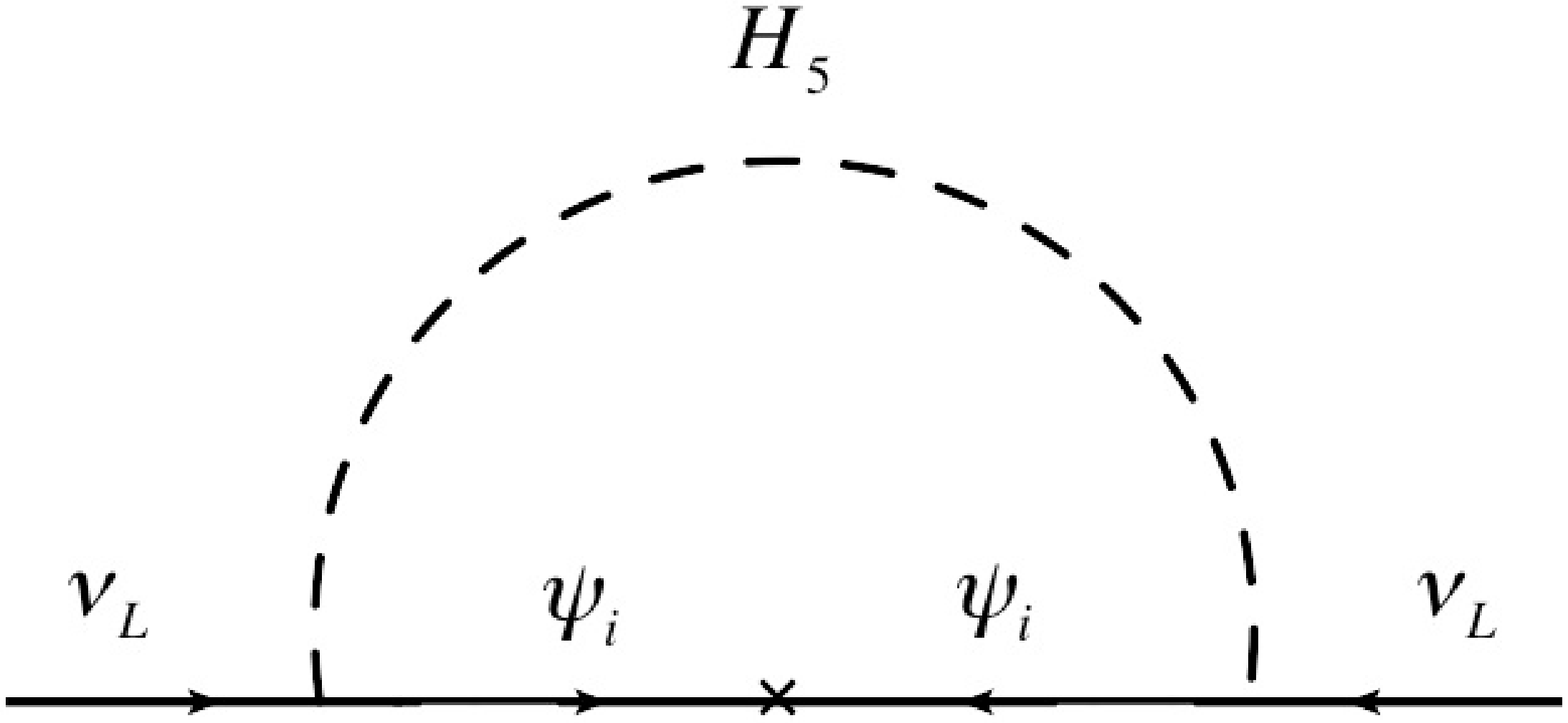}
\caption{The diagram inducing active neutrino mass.}   \label{fig:diagram}
\end{center}\end{figure}

%%%
\noindent \underline{\it  Active neutrino sector} :
 In our scenario, active neutrino mass is induced at one-loop level, where $\psi_{1,2}$ and $H_5$ propagate inside a loop diagram as in Fig.~\ref{fig:diagram},
 and the masses of real and imaginary part of electrically neutral component of $H_5$ are respectively denoted by $m_R$ and $m_I$.
As a result the active neutrino mass matrix is obtained such that
\begin{align}
m_\nu = \sum_{\alpha=1}^6 \frac{f_{i\alpha} M_\alpha f^T_{\alpha j} }{8(4\pi)^2}
\left[
\frac{r_R^\alpha \ln r^\alpha_R}{1-r^\alpha_R}
-
\frac{r_I^\alpha \ln r^\alpha_I}{1-r^\alpha_I}
\right],
\end{align}
where $r^{\alpha}_{R/I} \equiv\frac{m^2_{R/I}}{M_\alpha^2}$. 
Neutrino mass eigenvalues ($D_\nu$) are given by $D_\nu=U_{\rm MNS} m_\nu U^T_{\rm MNS}$, where $U_{\rm MNS}$ is  the MNS matrix.
Once we define $m_{\nu} \equiv f {\cal M} f^T$, one can rewrite $f$ in terms of the other parameters~\cite{Casas:2001sr, Chiang:2017tai} as follows:
\begin{align}
f_{ik}=\sum_{\alpha=1}^6 U^\dag_{ij} \sqrt{D_{\nu_{jj}}} O_{j\alpha} \sqrt{{\cal M}_{\alpha\alpha}} V^*_{\alpha k},
\end{align}
where $O$ is a three by six arbitrary matrix, satisfying $OO^T=1$, and $|f|\lesssim \sqrt{4\pi}$ is imposed not to exceed the perturbative limit.

%%%%%%%%%%%%%%%%%%%

 \subsection{Analysis of other phenomenological formulas}
\begin{figure}[tb]
\begin{center}
\includegraphics[width=10.0cm]{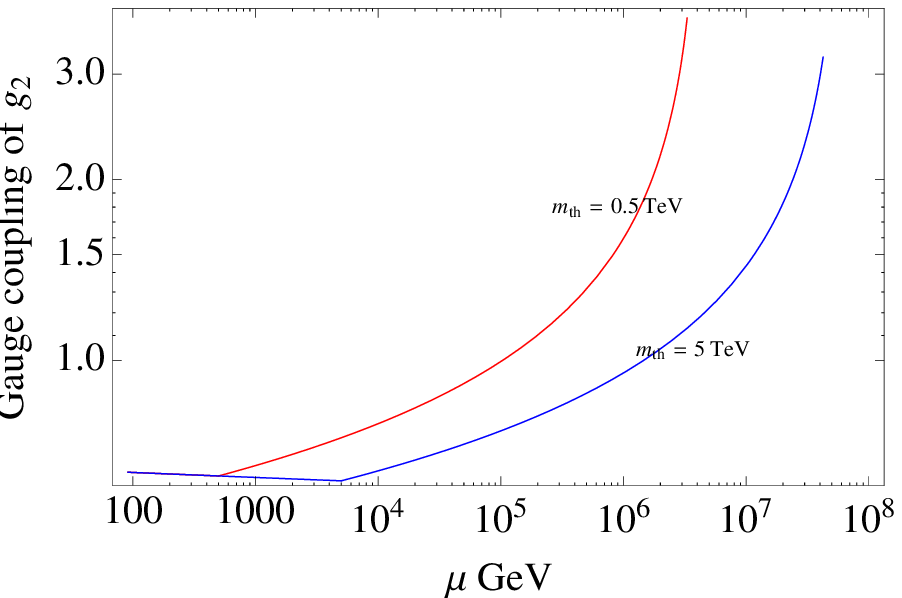}
\caption{The running of $g_2$ in terms of a reference energy of $\mu$, where the red line corresponds to  $m_{th}=$0.5 TeV,
while the blue one does  $m_{th}=$5 TeV. }
\label{fig:rge}
\end{center}\end{figure}

\noindent \underline{\it Beta function of $SU(2)_L$ gauge coupling $g_2$:}
\label{beta-func}
%%%
Here we estimate the running of gauge coupling of $g_2$ in the presence of several new multiplet fields of $SU(2)_L$.
The new contribution to $g_2$ from fermions (with three families) and  bosons are respectively given by~\cite{Nomura:2017abu, Kanemura:2015bli}
\begin{align}
 \Delta b^{f}_{g_2}=\frac{10}{3}, \ \Delta b^{b}_{g_2}=\frac{43}{3} .
\end{align}
%%%
Then one finds that the resulting flow of ${g_2}(\mu)$ is then given by the Fig.~\ref{fig:rge}.
This figure shows that the red line is relevant up to the mass scale $\mu={\cal O}(1)$ PeV in case of $m_{th}=$0.5 TeV,
while the blue line is relevant up to the mass scale $\mu={\cal O}(10)$ PeV in case of $m_{th}=$5 TeV.
%Thus our theory does not spoil, as far as we work on at around the scale of TeV.

%%%
\noindent \underline{\it  Lepton flavor violations(LFVs):}
LFV decays $\ell_i \to \ell_j \gamma$ arise from the term associated with coupling $f$ at one-loop level, and its form can be given by~\cite{Lindner:2016bgg, Baek:2016kud}
\begin{align}
 {\rm BR}(\ell_i\to\ell_j\gamma)= \frac{48\pi^3\alpha_{\rm em} C_{ij} }{{\rm G_F^2} m_{\ell_i}^2}\left(|a_{R_{ij}}|^2+|a_{L_{ij}}|^2\right),
 \end{align}
where 
 %%%
 \begin{align}
 a_{R_{ij}} &=\sum_{\alpha=1}^3
 \frac{f_{j\alpha} m_{\ell_i} f^\dag_{\alpha i}} {(4\pi)^2}
%\sum_{k=1,2,3}
\left[-\frac1{12} G(m_{a},M_{\pm_\alpha}) +G(M_\alpha,m_\pm)+G(M_{3+\alpha},m_\pm)
\right.\nn\\
&+ \left. \frac14\left[ 2 G(M_{\pm_\alpha}, m_{\pm\pm}) + G(m_{\pm\pm}, M_{\pm_\alpha}) \right]
- G(M_{\pm\pm_\alpha}, m_{\pm}) - 2 G(m_{\pm}, M_{\pm\pm_\alpha}) \right],\label{eq:amu}
\end{align}
and 
\begin{align}
&G(m_a,m_b)\equiv \int_0^1dx\int_0^{1-x}dy\frac{xy}{(x^2-x)m^2_{\ell_i} +x m_a^2+(1-x) m^2_b},
\end{align}
where $a_L=a_R(m_{\ell_i}\to m_{\ell_j})$.

%%%
\noindent \underline{\it New contributions to the muon anomalous magnetic moment} (muon $g-2$: $\Delta a_\mu$) :
We obtain $\Delta a_\mu$ from the same diagrams for LFVs and it can be formulated by the following expression
%%%
%Also another source via additional gauge sector can also be induced by
\begin{align}
&\Delta a_\mu \approx -m_\mu [{a_{L_{\mu\mu}}+a_{R_{\mu\mu}}}] 
= -2m_\mu{a_{L_{\mu\mu}}}, \label{eq:G2-ZP}
\end{align}
{where $a_{L_{\mu \mu}} = a_{R_{\mu \mu}}$ has been applied. In Eq.~(\ref{eq:amu}), one finds that the first term and the last two terms 
provide positive contributions, while the other terms do the negative contributions. When mediated masses are same value for all the modes;
$(m\equiv) m_a=m_\pm=m_{\pm\pm}=M_{\pm}=M_{\pm\pm}=M_{\pm\pm}$,
one simplifies the formula of $a_R$ as
   \begin{align}
 a_{R_{ij}} &\approx -\frac13\sum_{\alpha=1}^3
 \frac{f_{j\alpha} m_{\ell_i} f^\dag_{\alpha i}} {(4\pi)^2} G(m,m).
 \if0
 \approx 
 -\frac1{72}\sum_{\alpha=1}^3
 \frac{f_{j\alpha} m_{\ell_i} f^\dag_{\alpha i}} {(4\pi)^2m^2}. 
\fi\end{align}
Thus one would have positive contribution to the muon $g-2$,
and we use the allowed range of $\Delta a_\mu= (26.1\pm8.0)\times 10^{-10}$ in our numerical analysis below.
 %Also $M_{Z'}\lesssim$ 0.4 GeV has to be satisfied in addition to the constraint of  gauge coupling if only the pure gauge contribution is dominant.
}

{
\noindent \underline{\it Charged scalar contribution to $h \to \gamma \gamma$ decay}:
Interactions among SM Higgs field and large multiplet scalars affect the branching ratio of $h \to \gamma \gamma$ process via charged scalar loop.
Here we write the relevant interactions such that
\begin{equation}
\mathcal{V} \supset \sum_{\Phi = H_4, H_5, H_7} \lambda_{H \Phi} (H_2^\dagger H_2)(\Phi^\dagger \Phi) \supset  \sum_{\Phi = H_4, H_5, H_7} \lambda_{H \Phi} v_2 h (\Phi^\dagger \Phi),
\end{equation} 
where $\Phi^\dagger \Phi$ provide sum of charged scalar bilinear terms.
Then we obtain decay width of $h \to \gamma \gamma$ at one-loop level as~\cite{Gunion:1989we} 
\begin{equation}
\Gamma_{h \to \gamma \gamma} \simeq \frac{\alpha_{em}^2 m_h^3}{256 \pi^3} \left| \frac{4}{3 v_2} A_{1/2}(\tau_t) + \frac{1}{v_2} A_1(\tau_W) + \sum_\Phi \sum_{\Phi_i} Q_{\Phi_i}^2 \frac{\lambda_{H \Phi}}{2 m_\Phi^2} A_0(\tau_\Phi) \right|^2,
\end{equation}
where $\Phi_i$ indicates components in the multiplet $\Phi$ and $Q_{\Phi_i}$ is its electric charge, and $\tau_f = 4 m_f^2/m_h^2$.
The loop functions are given by
\begin{align}
& A_0 (x) = -x^2[x^{-1} - [\sin^{-1} (1/\sqrt{x})]^2], \\
& A_{1/2} (x) = 2 x^2[x^{-1} + (x^{-1} -1 )[\sin^{-1} (1/\sqrt{x})]^2], \\
& A_1 (x) = -x^2[2 x^{-2} + 3 x^{-1} + 3(2 x^{-1}-1) [\sin^{-1} (1/\sqrt{x})]^2]
\end{align}
where $x < 1$ is assumed and subscript of $A_{0,1/2,1}(x)$ correspond to spin of particle in loop diagram.
We then estimate $\mu_{\gamma \gamma} \equiv BR(h \to \gamma \gamma)_{\rm SM+ exotic}/BR(h \to \gamma \gamma)_{\rm SM}$ assuming Higgs production cross section is the same as in the SM.
In Fig.~\ref{fig:diphoton}, we show the $\mu_{\gamma \gamma}$ as a function of function of $\lambda_{H \Phi}$ assuming they are same value for $\Phi = (H_4, H_5, H_7)$ and masses of corresponding multiplets are $(1, 5, 1)$ TeV.
The value of $\mu_{\gamma \gamma}$ is constrained by the current LHC data~\cite{ATLAS:2018doi, Sirunyan:2018koj} and we indicate $1 \sigma$ region in the plot.
We thus find that $|\lambda_{H \Phi}|$ is required to be less than around $1$ for TeV scale scalar masses.

%%%%%%%%%%%%%%%%%%%
\begin{figure}[t]
\begin{center}
\includegraphics[width=10cm]{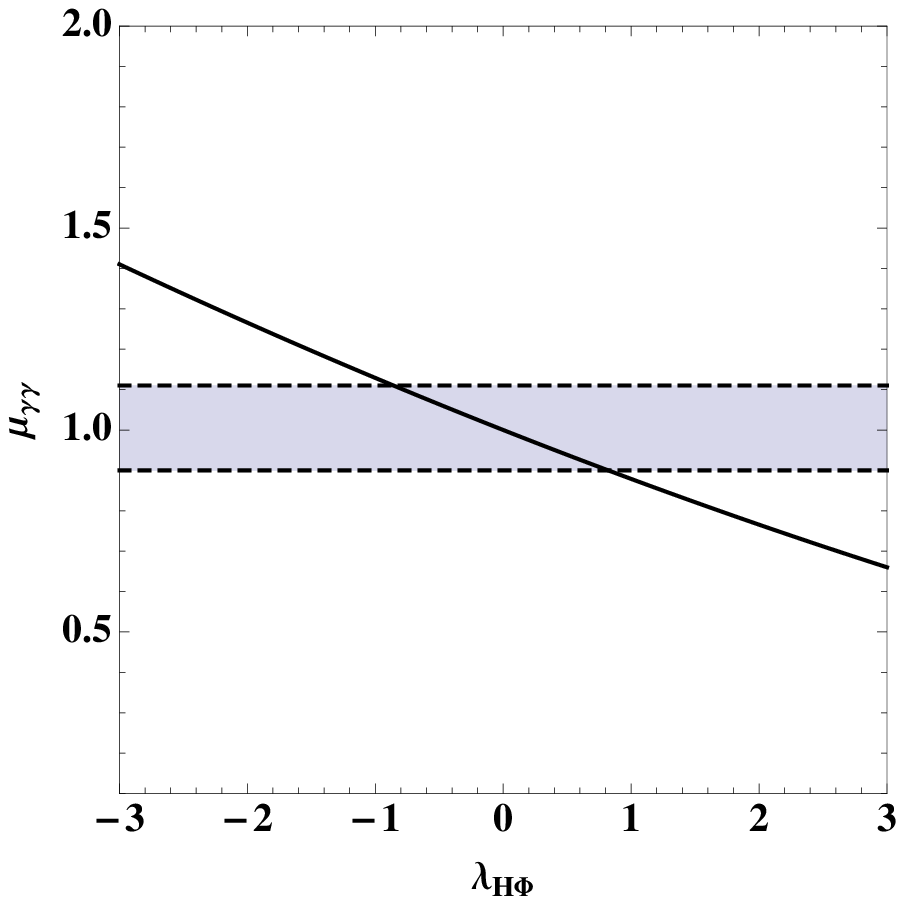}
\caption{$\mu_{\gamma \gamma} \equiv BR(h \to \gamma \gamma)_{\rm SM+ exotic}/BR(h \to \gamma \gamma)_{\rm SM}$ as a function of $\lambda_{H \Phi}$ assuming they are same value for $\Phi = H_4, H_5, H_7$ and masses of corresponding multiplets are $(1, 5, 1)$ TeV. The shaded region is $1 \sigma$ region from the LHC data~\cite{ATLAS:2018doi}.}   \label{fig:diphoton}
\end{center}\end{figure}
%%%%%%%%%%%%%%%%%%%

}

\noindent \underline{\it  Dark matter candidate}: 
In our case, the lightest neutral fermion among $\psi_{1,2}$ can be a DM candidate, which comes from $SU(2)_L$ quintet field with $-1/2$ charge under $U(1)_Y$.
{Here we firstly require that higher-dimensional operator inducing decay of the DM is not induced by the physics above cut-off scale so that decay of DM can only be induced via renormalizable Lagrangian in the model. 
Assuming the dominant contribution to explain the relic density originates from gauge interactions in the kinetic terms, the typical mass range is $M_{DM} \gtrsim 2.4$ TeV where
$M_{DM} = 2.4 \pm 0.06$ TeV is estimated by perturbative calculation~\cite{Cirelli:2005uq} and heavier mass is required including non-perturbative Sommerfeld enhancement effect~\cite{Cirelli:2007xd}.  
Then the typical order of spin independent cross section for DM-nucleon scattering via Z-portal is at around $1.6\times10^{-45}$ cm$^2$~\cite{Cirelli:2005uq} for $M_{DM} \simeq 2.4$ TeV, which marginally satisfies the current experimental data of direct detection searches such as LUX~\cite{Akerib:2016vxi}, XENON1T~\cite{Aprile:2017iyp}, and PandaX-II~\cite{Cui:2017nnn}; the direct detection constraint is weaker for heavier DM mass.
In the numerical analysis, below, we fix the DM mass to be 2.4 TeV as a reference value for simplicity. }
One feature of our model is possible instability of DM since we do not impose additional symmetry at TeV scale. 
We thus have to estimate the decay of DM so that the life time $\tau_{DM}=\Gamma^{-1}_{DM}$ does not exceed the age of universe that is around $4.35\times 10^{17}$ second.
The main decay channel arises from interactions associated with couplings $f$ and $\lambda_0$, when we neglect the effect of mixing among neutral bosons.
Then the three body decay ratio of $\Gamma(DM\to \nu_i h_{} h_{})$ via the neutral component of $H_5$ is given by
\begin{align}
\Gamma(DM\to\nu_i h_{}h_{})\approx \frac{\lambda_0^2 |f_{1i}|^2 M_{DM}^3 v_7^2 }{7680 m_R^4\pi^3}\lesssim 
 \frac{\lambda_0^2 |{\rm Max}[f_{1i}]|^2 M_{DM}^3 v_7^2 }{7680 m_R^4\pi^3},
\end{align}
where we assume the final states  to be massless, $m_R\approx m_I$, $M_{DM}$ is the mass of DM, and $h_{}$ is the SM Higgs.
In the numerical analysis, we will estimate the lifetime and show the allowed region, where we take the maximum value of $|f_{1a}|$.
~\footnote{In case where the neutral component of $H_5$ is DM candidate, $H_5$ decays into SM-like Higgs pairs via $\lambda_0$, and its decay rate is given by
$\frac{\lambda_0^2v_7^2}{800\pi M_X}$. Then the required lower bound of $\lambda_0$ is of the order $10^{-19}$ so that its lifetime is longer than the age of Universe, where DM is estimated as 5 TeV~\cite{Cirelli:2005uq}.}
%%%

\section{Numerical analysis and phenomenology}

Here we carry out numerical analysis to discuss consistency of our model under the constraints discussed in previous section.
Then we discuss collider physics focusing on charged scalar bosons in the model.

\noindent \underline{\it  Numerical analyses}: 
In our numerical analysis, we assume all the mass of $\psi_{1,2}$ to be the mass of DM; 2.4 TeV, and all the component of $H_5$ except $m_I$
to be degenerate, where $m_{I}=1.1 m_R$. These assumptions are reasonable in the aspect of oblique parameters in the multiplet fields~\cite{pdg}.
Also we  fix to be the following values so as to maximize the muon $g-2$:
\begin{align}
& O _{12}=0.895+12.3i ,\quad O _{23}=1.88+0.52i ,\quad O _{13}=0.4+0.6i,
\end{align}
where $O_{12,23,13}$ are arbitral mixing matrix with complex values that are introduced in the neutrino sector~\cite{Nomura:2018lsx, Chiang:2017tai}.
Notice here that we also impose $|f|\lesssim \sqrt{4\pi}$ not to exceed the perturbative limit.

\begin{figure}[t]
\begin{center}
\includegraphics[width=10cm]{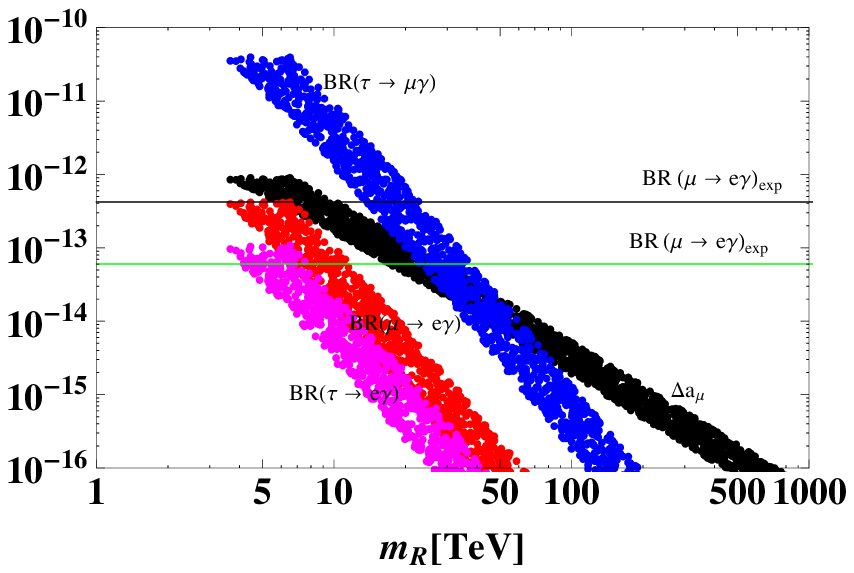}
\caption{Various LFV processes and $\Delta a_\mu$ in terms of  $m_R$, where $BR(\mu\to e\gamma)$, $BR(\tau\to e\gamma)$, $BR(\tau\to \mu\gamma)$, and $\Delta a_\mu$ are respectively colored by red, magenta, blue, and black.
The black horizontal line shows the current upper limit of the experiment~\cite{Cai:2017jrq, TheMEG:2016wtm}, while the green one does the future upper limit of the experiment~\cite{Cai:2017jrq, Baldini:2013ke}. }   \label{fig:f}
\end{center}\end{figure}

Fig.~\ref{fig:f}  represents various LFV processes and $\Delta a_\mu$ in terms of  $m_R$, where $BR(\mu\to e\gamma)$, $BR(\tau\to e\gamma)$, $BR(\tau\to \mu\gamma)$, and $\Delta a_\mu$ are respectively colored by red, magenta, blue, and black.
The black horizontal line shows the current upper limit of the experiment~\cite{Cai:2017jrq, TheMEG:2016wtm}, while the green one does the future upper limit of the experiment~\cite{Cai:2017jrq, Baldini:2013ke}. Considering these bounds of $\mu\to e\gamma$, one finds that the current allowed mass range of $m_R \sim$ 4-20 TeV can be tested in the near future.  
Here the upper bounds of $BR(\tau\to e\gamma)$ and $BR(\tau\to \mu\gamma)$ are of the order $10^{-8}$, which is safe for all the range. The maximum value of $\Delta a_\mu$ is about $10^{-12}$, which is smaller than the experimental value by three order of magnitude.

%%%%%%%%%%%%%%%%%%%
%%%%%%%%%%%%%%%%%%%

%%%%%%%%%%%%%%%%%%%%%

\begin{figure}[t]
\begin{center}
\includegraphics[width=10cm]{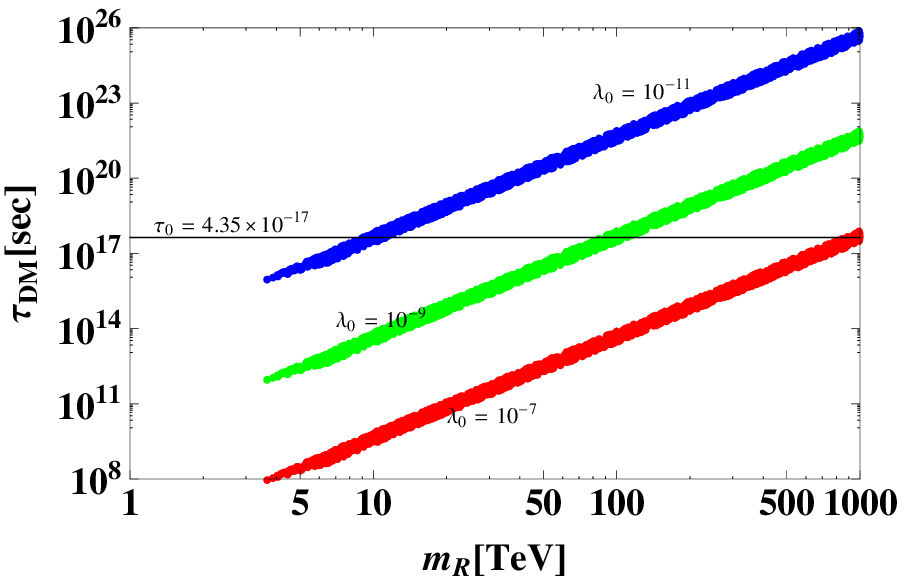}
\caption{ the lifetime of DM in terms of  $m_R$, where we fix $v_7\approx1.03$ GeV, and $\lambda_0=(10^{-7}, 10^{-9},10^{-11})$ with  
 (red, green, blue).
The black horizontal line shows the current age of Universe $\tau_0$. }   \label{fig:tau}
\end{center}\end{figure}

Fig.~\ref{fig:tau}  shows the lifetime of DM in terms of  $m_R$, where we fix $v_7\approx1.03$ GeV, and $\lambda_0=(10^{-7}, 10^{-9},10^{-11})$ with  
 (red, green, blue).
The black horizontal line shows the current age of Universe.
The figure demonstrates as follows:
\begin{align}
\lambda_0=10^{-7}:\ m_R\sim 1000\ {\rm TeV},\quad
\lambda_0=10^{-9}:\ 100\ {\rm TeV}\lesssim m_R,\quad
\lambda_0=10^{-11}:\ 10\ {\rm TeV}\lesssim m_R.
\end{align}

%%%%%%%%%%%%%%%%%%%
%%%%%%%%%%%%%%%%%%%

\begin{figure}[t]
\begin{center}
\includegraphics[width=10cm]{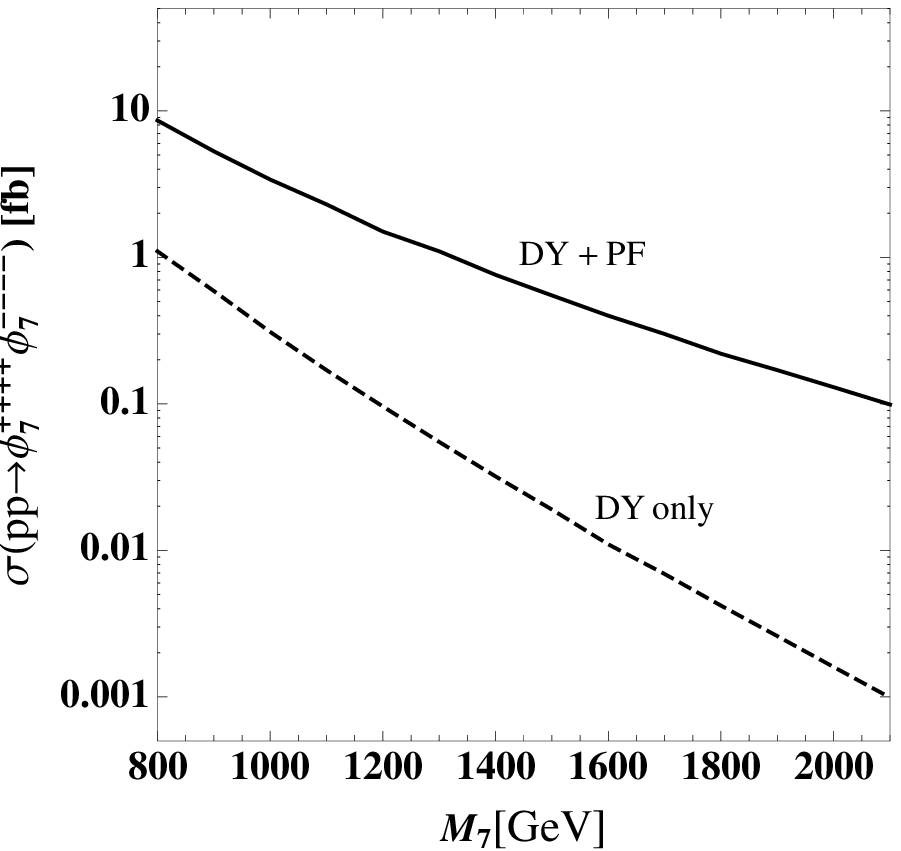}
\caption{Cross section for $pp \to \phi^{++++}_7 \phi^{----}_7$ at the LHC 14 TeV where dashed line indicate the cross section from only Drell-Yan process and solid line corresponds to the cross section including both Drell-Yan and photon fusion processes.}   \label{fig:LHC}
\end{center}\end{figure}

\noindent \underline{\it  Collider Physics}: 
Here let us briefly comments possible collider physics of our model.
We have many new charged particles from $SU(2)_L$ multiplet scalars and fermions.
Clear signal could be obtained from charged scalar bosons in $H_7$ and $H_4$, since they can decay into final states containing only SM particles
where the components in these multiplets are given by
\begin{align}
& H_7 = (\phi_7^{++++}, \phi_7^{+++}, \phi_7^{++}, \phi^{+}_7, \phi^{0}_7, \phi'^{-}_7, \phi'^{--})^T,  \label{eq:H7} \\
& H_4 = (\phi_4^{++}, \phi_4^{+}, \phi_4^{0}, \phi'^{-}_4)^T.  \label{eq:H4} 
\end{align}
The quadruply charged scalar is particularly interesting since it is specific in our model and would provide sizable production cross section.
We thus focus on $\phi_7^{\pm \pm \pm \pm}$ signal in our model~\footnote{Collider phenomenology of charged scalars from quartet is discussed in refs.~\cite{delAguila:2013yaa,delAguila:2013mia, Nomura:2017abu,Chala:2018ari}.}.
The quadruply charged scalar can be pair produced by Drell-Yan(DY) process, $q \bar q \to Z/\gamma \to \phi^{++++}_7 \phi^{----}_7$, and by photon fusion(PF) process $\gamma \gamma \to \phi^{++++}_7 \phi^{----}_7$~\cite{Babu:2016rcr, Ghosh:2017jbw, Ghosh:2018drw}. 
We estimate the cross section using {\tt MADGRAPH/MADEVENT\,5}~\cite{Alwall:2014hca}, where the necessary Feynman rules and relevant parameters of the model are implemented by use of FeynRules 2.0 \cite{Alloul:2013bka} and the {\tt NNPDF23LO1} PDF~\cite{Deans:2013mha} is adopted.
In Fig.~\ref{fig:LHC} we show the cross section for the quadruply charged scalar production process $pp \to \phi^{++++}_7 \phi^{----}_7$ at the LHC 14 TeV, where dashed line indicates the cross section from only Drell-Yan process and solid line corresponds to the cross section including both Drell-Yan and photon fusion processes.
We thus find that the cross section is highly enhanced including PF process due to large electric charge of the scalar boson.
Thus sizable number of $\phi_7^{\pm \pm \pm \pm}$ pair can be produced at the LHC 14 TeV if its mass is $\mathcal{O}(1)$ TeV, with sufficiently large integrated luminosity.
Produced $\phi^{\pm \pm \pm \pm}_7$ mainly decays into $\phi_4^{\pm \pm} \phi^{\pm \pm}_4$ via $H_4^T \tilde H_7 H_4$ interactions in the scalar potential since components in $H_7$ have degenerate mass.
Then $\phi_4^{\pm \pm}$ decays into $W^\pm W^\pm$ via $(D_\mu H_4)^\dagger (D^\mu H_4)$ term.
We thus obtain multi $W$ boson signal from quadruply charged scalar boson production.
Mass reconstruction from multi $W$ boson final state is not trivial and detailed analysis is beyond the scope of this paper.

%%%%%%%%%%%%%%%%%%%
%%%%%%%%%%%%%%%%%%%

\begin{figure}[t]
\begin{center}
\includegraphics[width=10cm]{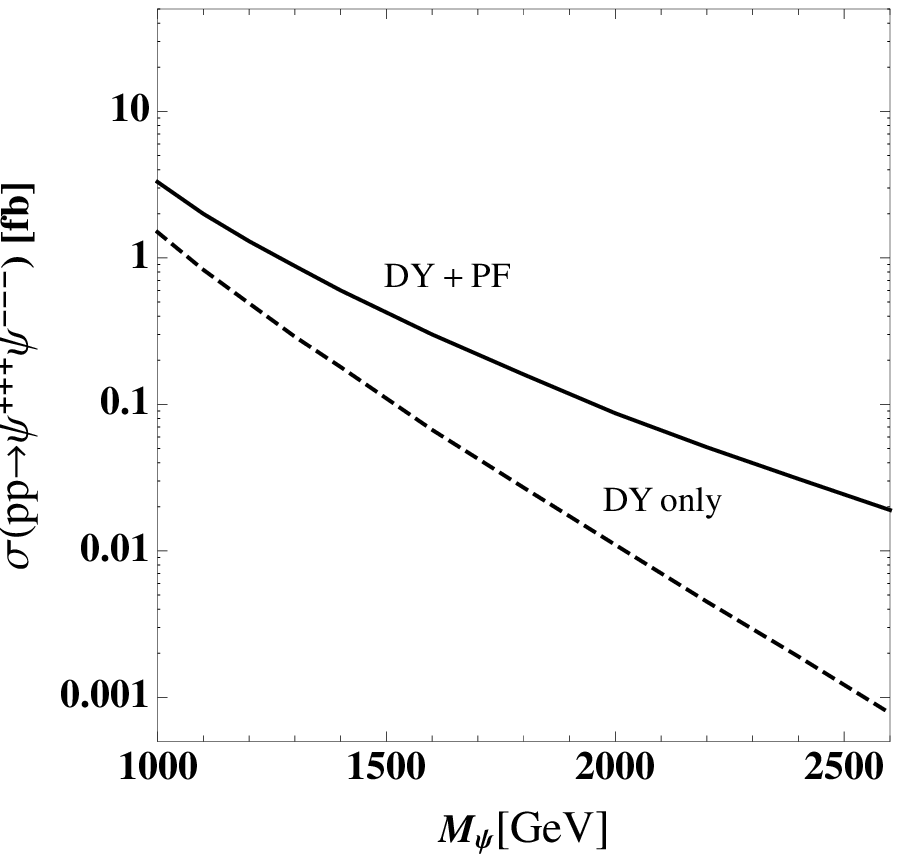}
\caption{Cross section for $pp \to \psi^{+++} \psi^{---}$ at the LHC 14 TeV where dashed line indicate the cross section from only Drell-Yan process and solid line corresponds to the cross section including both Drell-Yan and photon fusion processes.}   \label{fig:LHC2}
\end{center}\end{figure}

{ In addition to the charged scalar bosons, we consider production of exotic charged fermions at the LHC. 
The quadruplet fermion $\psi^a$ is written by 
\begin{equation}
\psi^a = (\psi^0, \psi^-, \psi^{--}, \psi^{---})^a
\end{equation}
where the subscript indicates electric charge of components.
As in the scalar sector, we focus on the component with the highest electric charge that is $\psi^{\pm \pm \pm}$ in the multiplet.
Pair production of $\psi^{\pm \pm \pm}$ is estimated by {\tt MADGRAPH/MADEVENT\,5} as in the charged scalar case 
where we consider both DY- and PF-processes. 
The production cross section is shown In Fig.~\ref{fig:LHC2} where the dashed and solid lines correspond to values from only DY process and from sum of both processes as in the scalar case.
We obtain cross section $\sigma \sim 0.03$ fb for $M_\psi \sim 2.4$ TeV which is motivated by DM relic density.
In that case we can obtain $\sim 10 (100)$ events for integrated luminosity of $300 (3000)$ fb.
Charged fermions in $\psi^a$ decay as $\psi^{n} \to \psi^{n\pm1} W^{\mp*}$ where $n$ indicates electric charge and $W$ boson is off-shell since 
the mass differences between components are radiatively induced and its value is around 350 MeV~\cite{Cirelli:2005uq}; exotic fermions cannot decay via $\bar L H_5 \psi$ coupling since $H_5$ is heavier than $\psi$.  
Thus $\psi^{\pm \pm \pm}$ production gives signature of light mesons with missing transverse momentum through decay chain of $\psi^{\pm \pm \pm} \to W^{\pm*} \psi^{\pm \pm} (\to W^{\pm*} \psi^\pm (\to W^{\pm*} \psi^0))$ where $\psi^0$ is DM. 
Furthermore we would have displaced vertex signature since decay length of charged fermions is long as $\mathcal{O}(1)$ cm~\cite{Cirelli:2005uq} for quadruplet fermion.
Therefore analysis of displaced vertex will be important to test our scenario. 
}

%%%%%%%%%%%%%%%%%%%
%%%%%%%%%%%%%%%%%%%

%%%%%%%%%%%%%%%%%%%%
\section{Summary and discussions}
We have proposed an one-loop neutrino mass model, introducing large multiplet fields under $SU(2)_L$.
The inert boson is achieved by nontrivial cancellations among quadratic terms.
We have also considered the RGE for $g_2$, the LFVs, muon $g-2$, and fermionic DM candidate,
and shown allowed region to satisfy all the constraints as we have discussed above.
RGE of $g_2$ determines our cut-off energy that does makes our theory stay within the order $10$ PeV scale,
therefore our model could totally be tested by current or near future experiments.    
%The fermionic DM could be more natural than the case of bosonic DM candidate, because we consider the lifetime.
Due to the multiplet fields, we have positive value of muon $g-2$, but find its maximum value to be of the order $10^{-12}$ that is smaller than the sizable value
by three order of magnitude. For the LFVs, the most promising mode to be tested in the current and future experiments is $\mu\to e\gamma$ at the range of 3.2 TeV $\lesssim m_R\lesssim$ 11 TeV. 
{We have also discussed possible decay mode of our DM candidate and some parameters are constrained requiring DM to be stable on cosmological time scale. Notice that the decay of DM is one feature of our model and we would discriminate our model from models with absolutely stable DM by searching for signal of the DM decay.}
Finally, we have analyzed the collider physics, focussing on multi-charged scalar bosons $H_4$ and $H_7$, {and triply charged fermion $\psi^{\pm \pm \pm}$ in exotic fermion sector}.
For scalar sector, we find that sizable production cross section for quadruply charged scalar pair can be obtained adding the photon fusion process that is enhanced by large electric charge of $\phi^{\pm\pm\pm\pm}_7$. 
Then possible signal of $\phi^{\pm\pm\pm\pm}_7$ comes from decay chain of $\phi^{\pm\pm\pm\pm}_7 \to \phi^{\pm\pm}_4 \phi^{\pm\pm}_4 \to 4 W^\pm$ which would provide multi-lepton plus jets at the detector. 
We expect sizable number of events with sufficiently large integrated luminosity to detect them at the LHC 14 TeV where the detailed analysis of the signal and background is left in future works. 
{For exotic fermion sector, we have also find sizable production cross section for triply charged fermion pair. 
The triply charged fermion decay gives signature of light mesons with missing transverse momentum through decay chain of $\psi^{\pm \pm \pm} \to W^{\pm*} \psi^{\pm \pm} (\to W^{\pm*} \psi^\pm (\to W^{\pm*} \psi^0))$ where $\psi^0$ is DM. 
In addition, would have displaced vertex signature since decay length of charged fermions is long as $\mathcal{O}(1)$ cm for components in quadruplet fermion, and
thus analysis of displaced vertex will be important to test our scenario.
}

%%%%%%%%%%%%%%%%%%%%%%%%%%%%%%%%%%%
\section*{Acknowledgments}
%\vspace{0.3cm}
This research is supported by the Ministry of Science, ICT and Future Planning, Gyeongsangbuk-do and Pohang City (H.O.). 
H. O. is sincerely grateful for KIAS and all the members.
%%%%%%%%%%%%%%%%%%%%%%%%%%%%%%%%%%%


\begin{thebibliography}{99}

%\cite{Ma:2006km}
\bibitem{Ma:2006km} 
  E.~Ma,
  %``Verifiable radiative seesaw mechanism of neutrino mass and dark matter,''
  Phys.\ Rev.\ D {\bf 73}, 077301 (2006)
  [hep-ph/0601225].
  %%CITATION = HEP-PH/0601225;%%
  %327 citations counted in INSPIRE as of 09 Dec 2013
  
   
  
%\cite{Krauss:2002px}
\bibitem{Krauss:2002px}
  L.~M.~Krauss, S.~Nasri and M.~Trodden,
  %``A Model for neutrino masses and dark matter,''
  Phys.\ Rev.\  D {\bf 67}, 085002 (2003)
  [arXiv:hep-ph/0210389].
  %%CITATION = PHRVA,D67,085002;%%


%\cite{Aoki:2008av}
\bibitem{Aoki:2008av}
  M.~Aoki, S.~Kanemura and O.~Seto,
  %``Neutrino mass, Dark Matter and Baryon Asymmetry via TeV-Scale
	%Physics
  %without Fine-Tuning,''
  Phys.\ Rev.\ Lett.\  {\bf 102}, 051805 (2009)
  [arXiv:0807.0361].
  %%CITATION = PRLTA,102,051805;%%
  
  %\cite{Gustafsson:2012vj}
\bibitem{Gustafsson:2012vj} 
  M.~Gustafsson, J.~M.~No and M.~A.~Rivera,
  %``Predictive Model for Radiatively Induced Neutrino Masses and Mixings with Dark Matter,''
  Phys.\ Rev.\ Lett.\  {\bf 110}, no. 21, 211802 (2013)
  Erratum: [Phys.\ Rev.\ Lett.\  {\bf 112}, no. 25, 259902 (2014)]
%  doi:10.1103/PhysRevLett.110.211802, 10.1103/PhysRevLett.112.259902
  [arXiv:1212.4806 [hep-ph]].
  %%CITATION = doi:10.1103/PhysRevLett.110.211802, 10.1103/PhysRevLett.112.259902;%%
  %127 citations counted in INSPIRE as of 08 Aug 2018

  
  %\cite{Okada:2012np}
\bibitem{Okada:2012np} 
  H.~Okada and T.~Toma,
  %``Fermionic Dark Matter in Radiative Inverse Seesaw Model with $U(1)_{B-L}$,''
  Phys.\ Rev.\ D {\bf 86}, 033011 (2012)
%  doi:10.1103/PhysRevD.86.033011
  [arXiv:1207.0864 [hep-ph]].
  %%CITATION = doi:10.1103/PhysRevD.86.033011;%%
  %53 citations counted in INSPIRE as of 15 Dec 2018
  
  
  %\cite{Kajiyama:2013zla}
\bibitem{Kajiyama:2013zla} 
  Y.~Kajiyama, H.~Okada and K.~Yagyu,
  %``Two Loop Radiative Seesaw Model with Inert Triplet Scalar Field,''
  Nucl.\ Phys.\ B {\bf 874}, 198 (2013)
%  doi:10.1016/j.nuclphysb.2013.05.020
  [arXiv:1303.3463 [hep-ph]].
  %%CITATION = doi:10.1016/j.nuclphysb.2013.05.020;%%
  %68 citations counted in INSPIRE as of 08 Aug 2018

%\cite{Kajiyama:2013rla}
\bibitem{Kajiyama:2013rla} 
  Y.~Kajiyama, H.~Okada and T.~Toma,
  %``Multicomponent dark matter particles in a two-loop neutrino model,''
  Phys.\ Rev.\ D {\bf 88}, no. 1, 015029 (2013)
%  doi:10.1103/PhysRevD.88.015029
  [arXiv:1303.7356 [hep-ph]].
  %%CITATION = doi:10.1103/PhysRevD.88.015029;%%
  %65 citations counted in INSPIRE as of 15 Dec 2018
  
  
  %\cite{Nomura:2018ktz}
\bibitem{Nomura:2018ktz} 
  T.~Nomura and H.~Okada,
  %``Inverse seesaw model with large $SU(2)_L$ multiplets and natural mass hierarchy,''
  arXiv:1809.06039 [hep-ph].
  %%CITATION = ARXIV:1809.06039;%%
  %1 citations counted in INSPIRE as of 15 Dec 2018
  
  %\cite{Nomura:2018ibs}
\bibitem{Nomura:2018ibs} 
  T.~Nomura and H.~Okada,
  %``A Linear seesaw model with hidden gauge symmetry,''
  arXiv:1806.07182 [hep-ph].
  %%CITATION = ARXIV:1806.07182;%%
  %4 citations counted in INSPIRE as of 15 Dec 2018
  
  %\cite{Nomura:2018lsx}
\bibitem{Nomura:2018lsx} 
  T.~Nomura and H.~Okada,
  %``One-loop neutrino mass model without any additional symmetries,''
  arXiv:1808.05476 [hep-ph].
  %%CITATION = ARXIV:1808.05476;%%
  %1 citations counted in INSPIRE as of 15 Dec 2018
    
  %\cite{Nomura:2018cle}
\bibitem{Nomura:2018cle} 
  T.~Nomura and H.~Okada,
  %``Neutrino mass generation with large $SU(2)_L$ multiplets under local $U(1)_{L_\mu - L_\tau}$ symmetry,''
  Phys.\ Lett.\ B {\bf 783}, 381 (2018)
%  doi:10.1016/j.physletb.2018.07.011
  [arXiv:1805.03942 [hep-ph]].
  %%CITATION = doi:10.1016/j.physletb.2018.07.011;%%
  %6 citations counted in INSPIRE as of 15 Dec 2018
  
  %\cite{Nomura:2018cfu}
\bibitem{Nomura:2018cfu} 
  T.~Nomura and H.~Okada,
  %``An inverse seesaw model with natural hierarchy at TeV scale,''
  arXiv:1807.04555 [hep-ph].
  %%CITATION = ARXIV:1807.04555;%%
  %3 citations counted in INSPIRE as of 15 Dec 2018
  
  %\cite{Nomura:2017abu}
\bibitem{Nomura:2017abu} 
  T.~Nomura and H.~Okada,
  %``Neutrino mass with large $SU(2)_L$ multiplet fields,''
  Phys.\ Rev.\ D {\bf 96}, no. 9, 095017 (2017)
%  doi:10.1103/PhysRevD.96.095017
  [arXiv:1708.03204 [hep-ph]].
  %%CITATION = doi:10.1103/PhysRevD.96.095017;%%
  %8 citations counted in INSPIRE as of 15 Dec 2018
  
  %\cite{Nomura:2016jnl}
\bibitem{Nomura:2016jnl} 
  T.~Nomura, H.~Okada and Y.~Orikasa,
  %``$SU(2)_L$ septet scalar linking to a radiative neutrino model,''
  Phys.\ Rev.\ D {\bf 94}, no. 5, 055012 (2016)
%  doi:10.1103/PhysRevD.94.055012
  [arXiv:1605.02601 [hep-ph]].
  %%CITATION = doi:10.1103/PhysRevD.94.055012;%%
  %25 citations counted in INSPIRE as of 15 Dec 2018
  
  
  %\cite{Anamiati:2018cuq}
\bibitem{Anamiati:2018cuq} 
  G.~Anamiati, O.~Castillo-Felisola, R.~M.~Fonseca, J.~C.~Helo and M.~Hirsch,
  %``High-dimensional neutrino masses,''
  arXiv:1806.07264 [hep-ph].
  %%CITATION = ARXIV:1806.07264;%%
  %1 citations counted in INSPIRE as of 13 Aug 2018
  

%\cite{Cirelli:2005uq}
\bibitem{Cirelli:2005uq} 
  M.~Cirelli, N.~Fornengo and A.~Strumia,
  %``Minimal dark matter,''
  Nucl.\ Phys.\ B {\bf 753}, 178 (2006)
%  doi:10.1016/j.nuclphysb.2006.07.012
  [hep-ph/0512090].
  %%CITATION = doi:10.1016/j.nuclphysb.2006.07.012;%%
  %571 citations counted in INSPIRE as of 09 Jul 2018
  
  %%%%%
  
  %%%%%%%%%%Neutrino Type-II %%%%%%%%%%%%%%
\bibitem{Magg:1980ut} 
  M.~Magg and C.~Wetterich,
  %``Neutrino Mass Problem and Gauge Hierarchy,''
  Phys.\ Lett.\ B {\bf 94}, 61 (1980);
  %%CITATION = PHLTA,B94,61;%%
  %457 citations counted in INSPIRE as of 30 Mar 2014
%  \bibitem{Lazarides:1980nt} 
  G.~Lazarides, Q.~Shafi and C.~Wetterich,
  %``Proton Lifetime and Fermion Masses in an SO(10) Model,''
  Nucl.\ Phys.\ B {\bf 181}, 287 (1981);
  %%CITATION = NUPHA,B181,287;%%
  %770 citations counted in INSPIRE as of 30 Mar 2014
%\bibitem{Mohapatra:1980yp} 
  R.~N.~Mohapatra and G.~Senjanovic,
  %``Neutrino Masses and Mixings in Gauge Models with Spontaneous Parity Violation,''
  Phys.\ Rev.\ D {\bf 23}, 165 (1981);
  %%CITATION = PHRVA,D23,165;%%
  %1610 citations counted in INSPIRE as of 30 Mar 2014
%\bibitem{Ma:1998dx} 
  E.~Ma and U.~Sarkar,
  %``Neutrino masses and leptogenesis with heavy Higgs triplets,''
  Phys.\ Rev.\ Lett.\  {\bf 80}, 5716 (1998)
  [hep-ph/9802445].
  %%CITATION = HEP-PH/9802445;%%
  %357 citations counted in INSPIRE as of 30 Mar 2014
  
 %%%%%%%%%%Neutrino Type-II further %%%%%%%%%%%
 \bibitem{Konetschny:1977bn} 
  W.~Konetschny and W.~Kummer,
  %``Nonconservation of Total Lepton Number with Scalar Bosons,''
  Phys.\ Lett.\ B {\bf 70}, 433 (1977);
  %%CITATION = PHLTA,B70,433;%%
  %178 citations counted in INSPIRE as of 30 Mar 2014
 %\bibitem{Schechter:1980gr} 
  J.~Schechter and J.~W.~F.~Valle,
  %``Neutrino Masses in SU(2) x U(1) Theories,''
  Phys.\ Rev.\ D {\bf 22}, 2227 (1980);
  %%CITATION = PHRVA,D22,2227;%%
  %1416 citations counted in INSPIRE as of 30 Mar 2014 
%\bibitem{Cheng:1980qt} 
  T.~P.~Cheng and L.~-F.~Li,
  %``Neutrino Masses, Mixings and Oscillations in SU(2) x U(1) Models of Electroweak Interactions,''
  Phys.\ Rev.\ D {\bf 22}, 2860 (1980);
  %%CITATION = PHRVA,D22,2860;%%
  %516 citations counted in INSPIRE as of 30 Mar 2014
%\bibitem{Bilenky:1980cx} 
  S.~M.~Bilenky, J.~Hosek and S.~T.~Petcov,
  %``On Oscillations of Neutrinos with Dirac and Majorana Masses,''
  Phys.\ Lett.\ B {\bf 94}, 495 (1980).
  %%CITATION = PHLTA,B94,495;%%
  %410 citations counted in INSPIRE as of 30 Mar 2014  
%%%%%%%%%%%%%%%%%%%%%%%%%%%%

  
  %%%%
  
  %\cite{Okada:2015bxa}
\bibitem{Okada:2015bxa} 
  H.~Okada, N.~Okada and Y.~Orikasa,
  %``Radiative seesaw mechanism in a minimal 3-3-1 model,''
  Phys.\ Rev.\ D {\bf 93}, no. 7, 073006 (2016)
%  doi:10.1103/PhysRevD.93.073006
  [arXiv:1504.01204 [hep-ph]].
  %%CITATION = doi:10.1103/PhysRevD.93.073006;%%
  %41 citations counted in INSPIRE as of 14 Dec 2018
      %%%%%%%%%%%%%%%%%%%%%%%%%%%%%%%%%%%%%%%

  
  
  %\cite{Patrignani:2016xqp}
\bibitem{pdg} 
  C.~Patrignani {\it et al.} [Particle Data Group],
  %``Review of Particle Physics,''
  Chin.\ Phys.\ C {\bf 40}, no. 10, 100001 (2016).
%  doi:10.1088/1674-1137/40/10/100001
  %%CITATION = doi:10.1088/1674-1137/40/10/100001;%%
  %4427 citations counted in INSPIRE as of 02 Dec 2018


%\cite{Casas:2001sr}
\bibitem{Casas:2001sr} 
  J.~A.~Casas and A.~Ibarra,
  %``Oscillating neutrinos and muon ---> e, gamma,''
  Nucl.\ Phys.\ B {\bf 618}, 171 (2001)
 % doi:10.1016/S0550-3213(01)00475-8
  [hep-ph/0103065].
  %%CITATION = doi:10.1016/S0550-3213(01)00475-8;%%
  %870 citations counted in INSPIRE as of 08 Aug 2018
  
%\cite{Chiang:2017tai}
\bibitem{Chiang:2017tai} 
  C.~W.~Chiang, H.~Okada and E.~Senaha,
  %``Dark matter, muon $g-2$, electric dipole moments, and $Z\to \ell_i^+ \ell_j^-$ in a one-loop induced neutrino model,''
  Phys.\ Rev.\ D {\bf 96}, no. 1, 015002 (2017)
%  doi:10.1103/PhysRevD.96.015002
  [arXiv:1703.09153 [hep-ph]].
  %%CITATION = doi:10.1103/PhysRevD.96.015002;%%
  %14 citations counted in INSPIRE as of 08 Aug 2018

  %\cite{Kanemura:2015bli}
\bibitem{Kanemura:2015bli} 
  S.~Kanemura, K.~Nishiwaki, H.~Okada, Y.~Orikasa, S.~C.~Park and R.~Watanabe,
  %``LHC 750 GeV diphoton excess in a radiative seesaw model,''
  PTEP {\bf 2016}, no. 12, 123B04 (2016)
%  doi:10.1093/ptep/ptw164
  [arXiv:1512.09048 [hep-ph]].
  %%CITATION = doi:10.1093/ptep/ptw164;%%
  %88 citations counted in INSPIRE as of 08 Jul 2018


%\cite{Lindner:2016bgg}
\bibitem{Lindner:2016bgg} 
  M.~Lindner, M.~Platscher and F.~S.~Queiroz,
  %``A Call for New Physics : The Muon Anomalous Magnetic Moment and Lepton Flavor Violation,''
  Phys.\ Rept.\  {\bf 731}, 1 (2018)
 % doi:10.1016/j.physrep.2017.12.001
  [arXiv:1610.06587 [hep-ph]].
  %%CITATION = doi:10.1016/j.physrep.2017.12.001;%%
  %84 citations counted in INSPIRE as of 08 Aug 2018

%\cite{Baek:2016kud}
\bibitem{Baek:2016kud} 
  S.~Baek, T.~Nomura and H.~Okada,
  %``An explanation of one-loop induced h → μτ decay,''
  Phys.\ Lett.\ B {\bf 759}, 91 (2016)
 % doi:10.1016/j.physletb.2016.05.055
  [arXiv:1604.03738 [hep-ph]].
  %%CITATION = doi:10.1016/j.physletb.2016.05.055;%%
  %18 citations counted in INSPIRE as of 08 Aug 2018


   \bibitem{Gunion:1989we} 
  J.~F.~Gunion, H.~E.~Haber, G.~L.~Kane and S.~Dawson,
  %``The Higgs Hunter's Guide,''
  Front.\ Phys.\  {\bf 80}, 1 (2000).
  %%CITATION = FRPHA,80,1;%%
  %396 citations counted in INSPIRE as of 12 Jul 2015


%\cite{ATLAS:2018doi}
\bibitem{ATLAS:2018doi} 
  The ATLAS collaboration [ATLAS Collaboration],
  %``Combined measurements of Higgs boson production and decay using up to 80 fb$^{-1}$ of proton--proton collision data at $\sqrt{s}=$ 13 TeV collected with the ATLAS experiment,''
  ATLAS-CONF-2018-031.
  %%CITATION = ATLAS-CONF-2018-031;%%
  %35 citations counted in INSPIRE as of 09 Feb 2019
  
  %\cite{Sirunyan:2018koj}
\bibitem{Sirunyan:2018koj} 
  A.~M.~Sirunyan {\it et al.} [CMS Collaboration],
  %``Combined measurements of Higgs boson couplings in proton-proton collisions at $\sqrt{s}=$ 13 TeV,''
  %Submitted to: Eur.Phys.J.
  [arXiv:1809.10733 [hep-ex]].
  %%CITATION = ARXIV:1809.10733;%%
  %33 citations counted in INSPIRE as of 09 Feb 2019


%\cite{Cirelli:2007xd}
\bibitem{Cirelli:2007xd} 
  M.~Cirelli, A.~Strumia and M.~Tamburini,
  %``Cosmology and Astrophysics of Minimal Dark Matter,''
  Nucl.\ Phys.\ B {\bf 787}, 152 (2007)
%  doi:10.1016/j.nuclphysb.2007.07.023
  [arXiv:0706.4071 [hep-ph]].
  %%CITATION = doi:10.1016/j.nuclphysb.2007.07.023;%%
  %395 citations counted in INSPIRE as of 10 Aug 2018 
  
    
  
  %\cite{Akerib:2016vxi}
\bibitem{Akerib:2016vxi} 
  D.~S.~Akerib {\it et al.} [LUX Collaboration],
  %``Results from a search for dark matter in the complete LUX exposure,''
  Phys.\ Rev.\ Lett.\  {\bf 118}, no. 2, 021303 (2017)
%  doi:10.1103/PhysRevLett.118.021303
  [arXiv:1608.07648 [astro-ph.CO]].
  %%CITATION = doi:10.1103/PhysRevLett.118.021303;%%
  %754 citations counted in INSPIRE as of 08 Oct 2018

%\cite{Aprile:2017iyp}
\bibitem{Aprile:2017iyp} 
  E.~Aprile {\it et al.} [XENON Collaboration],
  %``First Dark Matter Search Results from the XENON1T Experiment,''
  Phys.\ Rev.\ Lett.\  {\bf 119}, no. 18, 181301 (2017)
%  doi:10.1103/PhysRevLett.119.181301
  [arXiv:1705.06655 [astro-ph.CO]].
  %%CITATION = doi:10.1103/PhysRevLett.119.181301;%%
  %480 citations counted in INSPIRE as of 08 Oct 2018
  
  %\cite{Cui:2017nnn}
\bibitem{Cui:2017nnn} 
  X.~Cui {\it et al.} [PandaX-II Collaboration],
  %``Dark Matter Results From 54-Ton-Day Exposure of PandaX-II Experiment,''
  Phys.\ Rev.\ Lett.\  {\bf 119}, no. 18, 181302 (2017)
%  doi:10.1103/PhysRevLett.119.181302
  [arXiv:1708.06917 [astro-ph.CO]].
  %%CITATION = doi:10.1103/PhysRevLett.119.181302;%%
  %235 citations counted in INSPIRE as of 08 Oct 2018
  
  %\cite{Cai:2017jrq}
\bibitem{Cai:2017jrq} 
  Y.~Cai, J.~Herrero-Garcia, M.~A.~Schmidt, A.~Vicente and R.~R.~Volkas,
  %``From the trees to the forest: a review of radiative neutrino mass models,''
  Front.\ in Phys.\  {\bf 5}, 63 (2017)
%  doi:10.3389/fphy.2017.00063
  [arXiv:1706.08524 [hep-ph]].
  %%CITATION = doi:10.3389/fphy.2017.00063;%%
  %39 citations counted in INSPIRE as of 02 Dec 2018
    
%\cite{TheMEG:2016wtm}
\bibitem{TheMEG:2016wtm} 
  A.~M.~Baldini {\it et al.} [MEG Collaboration],
  %``Search for the lepton flavour violating decay $\mu ^+ \rightarrow \mathrm {e}^+ \gamma $ with the full dataset of the MEG experiment,''
  Eur.\ Phys.\ J.\ C {\bf 76}, no. 8, 434 (2016)
%  doi:10.1140/epjc/s10052-016-4271-x
  [arXiv:1605.05081 [hep-ex]].
  %%CITATION = doi:10.1140/epjc/s10052-016-4271-x;%%
  %247 citations counted in INSPIRE as of 02 Dec 2018

  %\cite{Baldini:2013ke}
\bibitem{Baldini:2013ke} 
  A.~M.~Baldini {\it et al.},
  %``MEG Upgrade Proposal,''
  arXiv:1301.7225 [physics.ins-det].
  %%CITATION = ARXIV:1301.7225;%%
  %251 citations counted in INSPIRE as of 02 Dec 2018


  %\cite{delAguila:2013yaa}
\bibitem{delAguila:2013yaa} 
  F.~del Aguila, M.~Chala, A.~Santamaria and J.~Wudka,
  %``Discriminating between lepton number violating scalars using events with four and three charged leptons at the LHC,''
  Phys.\ Lett.\ B {\bf 725}, 310 (2013)
%  doi:10.1016/j.physletb.2013.07.014
  [arXiv:1305.3904 [hep-ph]].
  %%CITATION = doi:10.1016/j.physletb.2013.07.014;%%
  %21 citations counted in INSPIRE as of 29 May 2018
  
%\cite{delAguila:2013mia}
\bibitem{delAguila:2013mia} 
  F.~del \'Aguila and M.~Chala,
  %``LHC bounds on Lepton Number Violation mediated by doubly and singly-charged scalars,''
  JHEP {\bf 1403}, 027 (2014)
%  doi:10.1007/JHEP03(2014)027
  [arXiv:1311.1510 [hep-ph]].
  %%CITATION = doi:10.1007/JHEP03(2014)027;%%
  %33 citations counted in INSPIRE as of 29 May 2018

%\cite{Chala:2018ari}
\bibitem{Chala:2018ari} 
  M.~Chala, C.~Krause and G.~Nardini,
  %``Signals of the electroweak phase transition at colliders and gravitational wave observatories,''
  arXiv:1802.02168 [hep-ph].
  %%CITATION = ARXIV:1802.02168;%%
  %2 citations counted in INSPIRE as of 29 May 2018
  

  
  %\cite{Babu:2016rcr}
\bibitem{Babu:2016rcr} 
  K.~S.~Babu and S.~Jana,
  %``Probing Doubly Charged Higgs Bosons at the LHC through Photon Initiated Processes,''
  Phys.\ Rev.\ D {\bf 95}, no. 5, 055020 (2017)
%  doi:10.1103/PhysRevD.95.055020
  [arXiv:1612.09224 [hep-ph]].
  %%CITATION = doi:10.1103/PhysRevD.95.055020;%%
  %26 citations counted in INSPIRE as of 11 Dec 2018
  
  %\cite{Ghosh:2017jbw}
\bibitem{Ghosh:2017jbw} 
  K.~Ghosh, S.~Jana and S.~Nandi,
  %``Neutrino Mass Generation at TeV Scale and New Physics Signatures from Charged Higgs at the LHC for Photon Initiated Processes,''
  JHEP {\bf 1803}, 180 (2018)
%  doi:10.1007/JHEP03(2018)180
  [arXiv:1705.01121 [hep-ph]].
  %%CITATION = doi:10.1007/JHEP03(2018)180;%%
  %6 citations counted in INSPIRE as of 11 Dec 2018
  
  %\cite{Ghosh:2018drw}
\bibitem{Ghosh:2018drw} 
  T.~Ghosh, S.~Jana and S.~Nandi,
  %``Neutrino mass from Higgs quadruplet and multicharged Higgs searches at the LHC,''
  Phys.\ Rev.\ D {\bf 97}, no. 11, 115037 (2018)
%  doi:10.1103/PhysRevD.97.115037
  [arXiv:1802.09251 [hep-ph]].
  %%CITATION = doi:10.1103/PhysRevD.97.115037;%%
  %2 citations counted in INSPIRE as of 11 Dec 2018
  
  %\cite{Alwall:2014hca}
\bibitem{Alwall:2014hca} 
  J.~Alwall {\it et al.},
  %``The automated computation of tree-level and next-to-leading order differential cross sections, and their matching to parton shower simulations,''
  JHEP {\bf 1407}, 079 (2014)
  %doi:10.1007/JHEP07(2014)079
  [arXiv:1405.0301 [hep-ph]].
  %%CITATION = doi:10.1007/JHEP07(2014)079;%%
  %960 citations counted in INSPIRE as of 21 May 2016

  
  \bibitem{Alloul:2013bka} 
  A.~Alloul, N.~D.~Christensen, C.~Degrande, C.~Duhr and B.~Fuks,
  %``FeynRules  2.0 - A complete toolbox for tree-level phenomenology,''
  Comput.\ Phys.\ Commun.\  {\bf 185}, 2250 (2014)
  [arXiv:1310.1921 [hep-ph]].
  %%CITATION = ARXIV:1310.1921;%%
  %139 citations counted in INSPIRE as of 07 Jan 2015
  
  %\cite{Deans:2013mha}
\bibitem{Deans:2013mha} 
  C.~S.~Deans [NNPDF Collaboration],
  %``Progress in the NNPDF global analysis,''
  arXiv:1304.2781 [hep-ph].
  %%CITATION = ARXIV:1304.2781;%%
  %1 citations counted in INSPIRE as of 07 Jan 2015




\if0
%\cite{Cheung:2016fjo}
\bibitem{Cheung:2016fjo} 
  K.~Cheung, T.~Nomura and H.~Okada,
  %``Testable radiative neutrino mass model without additional symmetries and explanation for the $b \to s \ell^+ \ell^-$ anomaly,''
  Phys.\ Rev.\ D {\bf 94}, no. 11, 115024 (2016)
%  doi:10.1103/PhysRevD.94.115024
  [arXiv:1610.02322 [hep-ph]].
  %%CITATION = doi:10.1103/PhysRevD.94.115024;%%
  %19 citations counted in INSPIRE as of 08 Aug 2018
  
  %\cite{Nomura:2016ask}
\bibitem{Nomura:2016ask} 
  T.~Nomura and H.~Okada,
  %``An Extended Colored Zee-Babu Model,''
  Phys.\ Rev.\ D {\bf 94}, 075021 (2016)
%  doi:10.1103/PhysRevD.94.075021
  [arXiv:1607.04952 [hep-ph]].
  %%CITATION = doi:10.1103/PhysRevD.94.075021;%%
  %18 citations counted in INSPIRE as of 08 Aug 2018

%\cite{Cheung:2018itc}
\bibitem{Cheung:2018itc} 
  K.~Cheung and H.~Okada,
  %``Generalized one-loop neutrino mass model with charged particles,''
  Phys.\ Rev.\ D {\bf 97}, no. 7, 075027 (2018)
 %  doi:10.1103/PhysRevD.97.075027
  [arXiv:1801.00585 [hep-ph]].
  %%CITATION = doi:10.1103/PhysRevD.97.075027;%%
  %1 citations counted in INSPIRE as of 08 Aug 2018
  
  %\cite{Cheung:2017kxb}
\bibitem{Cheung:2017kxb} 
  K.~Cheung and H.~Okada,
  %``A testable radiative neutrino mass model with multi-charged particles,''
  Phys.\ Lett.\ B {\bf 774}, 446 (2017)
%  doi:10.1016/j.physletb.2017.10.010
  [arXiv:1708.06111 [hep-ph]].
  %%CITATION = doi:10.1016/j.physletb.2017.10.010;%%
  %1 citations counted in INSPIRE as of 08 Aug 2018


%\cite{Peskin:1990zt}
\bibitem{Peskin:1990zt} 
  M.~E.~Peskin and T.~Takeuchi,
  %``A New constraint on a strongly interacting Higgs sector,''
  Phys.\ Rev.\ Lett.\  {\bf 65}, 964 (1990).
 % doi:10.1103/PhysRevLett.65.964
  %%CITATION = doi:10.1103/PhysRevLett.65.964;%%
  %1765 citations counted in INSPIRE as of 08 Aug 2018

%\cite{Peskin:1991sw}
\bibitem{Peskin:1991sw} 
  M.~E.~Peskin and T.~Takeuchi,
  %``Estimation of oblique electroweak corrections,''
  Phys.\ Rev.\ D {\bf 46}, 381 (1992).
 % doi:10.1103/PhysRevD.46.381
  %%CITATION = doi:10.1103/PhysRevD.46.381;%%
  %1985 citations counted in INSPIRE as of 08 Aug 2018

%\cite{Nomura:2017ezy}
\bibitem{Nomura:2017ezy} 
  T.~Nomura and H.~Okada,
  %``Loop suppressed light fermion masses with $U(1)_R$ gauge symmetry,''
  Phys.\ Rev.\ D {\bf 96}, no. 1, 015016 (2017)
%  doi:10.1103/PhysRevD.96.015016
  [arXiv:1704.03382 [hep-ph]].
  %%CITATION = doi:10.1103/PhysRevD.96.015016;%%
  %10 citations counted in INSPIRE as of 08 Aug 2018
  \fi
 
\end{thebibliography}
\end{document}